# Soft or Stiff?

## Stroop Tasks in Visuo-Tactile Tapping Interactions

Ryotaro Ishikawa and Taku Hachisu

*University of Tsukuba, Ibaraki, Japan*

(Email: {Ishikawa, hachisu}@ah.iit.tsukuba.ac.jp)

**Abstract** --- One of the key challenges in the field of haptic research is designing plausible stimuli using haptic interfaces with limited degrees of freedom. Although the plausible approach, which simplifies and/or exaggerates stimuli to enhance information transfer or create an artistic effect, has proven effective, evaluations of such stimuli have traditionally relied on subjective measures. This study aims to establish an objective evaluation method for haptic stimuli designed using the plausible approach. Focusing on stiffness/material perception, we developed a Stroop test within visuo-tactile tapping interactions in a virtual space. The demonstration system presents visual (textures) and tactile (vibration) stimuli at the moment of contact between a stylus and a cube, prompting participants to immediately identify the material they perceive visually. If the tactile stimuli are perceived as plausible, reaction times will be longer when the visual and tactile stimuli represent different materials than when they represent the same material.

Keywords: plausible rendering, Stroop effect, tapping, vibrotactile

## 1 INTRODUCTION

One of the key challenges in the field of haptic research is designing plausible stimuli using haptic interfaces with limited degrees of freedom. Here, we consider the context of tapping objects with a tool like a stylus, where humans can perceive the stiffness and material of objects through the resulting vibrations from contact [1]. To artificially reproduce authentic vibrotactile stimuli, the dynamics of the physical phenomenon and the characteristics of the haptic interface must be accurately reflected in the vibrotactile signal design [1]. In contrast, the plausible approach simplifies and/or exaggerates stimuli to enhance information transfer or create an artistic effect [3]. For instance, Okamura et al. modified vibrotactile signals based on human perceptual characteristics [4].

Evaluating the plausibility of such stimuli is crucial because assessing physical similarity is unfeasible when the modified signals deviate from physical accuracy. Consequently, previous research has relied on subjective evaluations, such as Likert scales and classification tasks, despite their susceptibility to subjective bias [5].

This study aims to establish an objective evaluation method for haptic stimuli designed using the plausible approach. In the context of stiffness and material perception, we designed a Stroop test [6] within visuo-tactile tapping interactions in a virtual space. The demonstration system presents visual (textures) and tactile (vibration) stimuli at the moment of contact between a stylus and a cube, prompting participants to immediately identify the material they perceive visually (see Fig. 1 ). If the tactile stimuli are perceived as plausible, reaction times will be longer when the visual and tactile stimuli represent different materials compared to when they represent the same material.

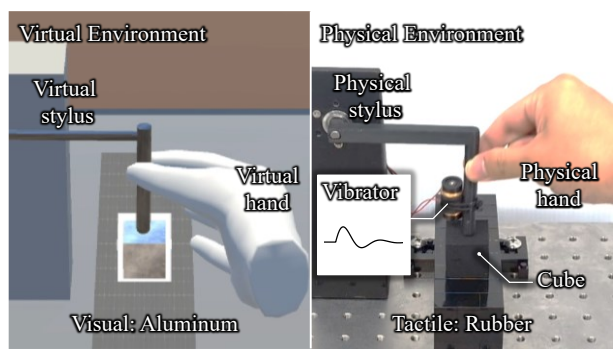

Fig.1 Stroop tasks in visuo-tactile tapping interactions: The system presents visual stimuli (textures) in a virtual space and tactile (vibration) stimuli in a physical space at the moment of contact between a stylus and a cube, prompting participants to immediately identify the material they perceive visually.

## 2 DEMONSTRATION SYSTEM

The apparatus comprises a haptic interface, visual display (either a head-mounted display (HMD, Meta Quest 2) or a monitor), keypad, headphones, and a host computer (ASUS, ROG Strix SCAR 16 G634JZ).

The haptic interface consists of a stylus equipped with a rotary encoder (OMRON, E6B2-CWZ6C, resolution: 2000P/R) and a vibrator (Tactile Lab, TL-002-14R Haptuator Redesign), a 30-mm foam cube, and a control circuit. The stylus is adjusted for perpendicular contact with the cube surface. A microcontroller (Espressif Systems, ESP32 DevKitC) runs a 10-kHz loop to read the encoder, detects contact between the stylus tip and the cubes, and computes the contact velocity. Upon contact detection, the microcontroller outputs decaying sinusoidal signals via a digital-to-analog converter (Microchip Technology/Atmel, MCP4922-E/P, resolution: 12 bits) and an audio amplifier (North Flat Japan, FX202A/FX-36A PRO) to drive the vibrator:

$$V(t) = Av \exp(-Bt) \sin(2\pi f t) \qquad (1)$$

where $A$, $B$, $f$, $v$, and $t$ are the initial amplitude coefficient, decay rate, frequency, impact velocity, and time, respectively. This demonstration system employs Okamura's model [4] to present plausible vibrotactile stimuli representing rubber or aluminum.

The computer renders a three-dimensional virtual space using Unity via the display. The haptic interface is placed similarly in both the virtual and physical spaces. A virtual 30-mm cube is initially rendered as a wireframe. Upon contact detection, it is rendered with a texture representing rubber or aluminum.

The headphones are used to present white noise to mask auditory cues, while the keypad is used to record participants' responses (the material they perceive visually).

## 3 DEMONSTRATION

Participants experience the visuo-tactile Stroop effect, in which they feel their slower and/or less accurate responses when visual and tactile stimuli represent incongruent materials than when they are congruent. Participants are instructed to be seated in front of the apparatus, hold a stylus-like pen, and tap the cube at a speed of approximately 1 m/s or less so as not to exceed the amplifier's current limit. They are also instructed to answer the material they perceive visually by using the keypad immediately after the contact between the stylus tip and the cube.

First, participants are asked to practice under unimodal conditions to familiarize themselves with the apparatus and procedures. In this phase, only visual stimuli are presented while tactile stimuli are absent (the vibrator is not driven). Then, they are asked to perform tasks under both congruent and incongruent conditions. Each condition consists of six trials, with the materials (rubber or aluminum) being randomly presented. In the congruent condition, visual and tactile stimuli represent the same material. In the incongruent condition, visual and tactile stimuli represent different materials. Finally, the difference in average reaction times between these conditions is displayed. The expected demonstration time per participant is a few minutes.

### ACKNOWLEDGEMENT

This study was funded by JST CREST (grant number JPMJCR22P4), Japan.